\documentclass[preprint,aps,amsmath,nofootinbib,amssymb,showkeys]{revtex4-1}
\usepackage{CJK,upgreek,fancyhdr}
\usepackage{verbatim,graphics,graphicx,color,slashed,textcomp,bm}
\usepackage{subcaption}
\usepackage[normalem]{ulem} 
\usepackage[colorlinks=true,linkcolor=red,urlcolor=blue,citecolor=blue]{hyperref}

\graphicspath{{figures/}}

\newcommand{\GeV}{\ \mathrm{GeV}}
\newcommand{\TeV}{\ \mathrm{TeV}}

\begin{document}

\title{Flavor Non-universal Gauge Interactions \\ and Anomalies in B-Meson Decays}

\author{Yong Tang$^{1}$ and Yue-Liang Wu$^{2,3,4}$}
\affiliation{
	${}^1$Department of Physics, Faculty of Science, \\
	The University of Tokyo, Bunkyo-ku, Tokyo 113-0033, Japan\\
	${}^2$International Centre for Theoretical Physics Asia-Pacific, Beijing, China \\
	${}^3$Institute of Theoretical Physics, Chinese Academy of Sciences, Beijing 100190, China\\
	${}^4$University of Chinese Academy of Sciences, Beijing 100049, China}

\begin{abstract}
Motivated by flavor non-universality and anomalies in semi-leptonic B-meson decays, we present a general and systematic discussion about how to construct anomaly-free $U(1)'$ gauge theories based on an extended standard model with only three right-handed neutrinos. If all standard model fermions are vector-like under this new gauge symmetry, the most general family non-universal charge assignments, $(a,b,c)$ for three-generation quarks and $(d,e,f)$ for leptons, need satisfy just one condition to be anomaly-free, $3(a+b+c)=-(d+e+f)$. Any assignment can be linear combinations of five independent anomaly-free solutions. We also illustrate how such models can generally lead to flavor-changing interactions and easily resolve the anomalies in B-meson decays. Probes with $B_s-\bar{B}_s$ mixing, decay into $\tau^\pm$, dilepton and dijet searches at colliders are also discussed.

\end{abstract}


\maketitle
\section{Introduction}
Recently, some intriguing anomalies have been found in the branching ratios of semi-leptonic decays of B-mesons  into electron and muon pairs, 
\begin{equation}\label{eq:rk}
\mathcal{R}_{K}=\frac{Br\left(B^+\rightarrow K^+\mu^+\mu^-\right)}{Br\left(B^+\rightarrow K^+ e^+e^-\right)},\;
\mathcal{R}_{K^*}=\frac{Br\left(B\rightarrow K^{*}\mu^+\mu^-\right)}{Br\left(B\rightarrow K^{*}e^+e^-\right)}. 
\end{equation}
The LHCb collaboration presented  the following values in Refs.~\cite{Aaij:2014ora, Aaij:2017vbb, RKstar}:
\begin{align}
\mathcal{R}_{K}   &= 0.745^{+0.090}_{-0.074}({\rm stat}) \pm 0.036({\rm syst}), \\
\mathcal{R}_{K^*} &= \left\{
\begin{array}{ll}
0.66^{+0.11}_{-0.07} \pm 0.03, & \textrm{for } (2m_\mu)^2 < q^2 < 1.1 \GeV^2,  \\[2mm]
0.69^{+0.11}_{-0.07} \pm 0.05, & \textrm{for } 1.1 \GeV^2 < q^2 < 6.0 \GeV^2,
\end{array}
\right.
\end{align}
where $q^2$ is the invariant mass for the final lepton pair. However, the standard model (SM) predicts $\mathcal{R}_{K}\approx 1\approx\mathcal{R}_{K^*}$~\cite{Hiller:2003js, Bordone:2016gaq} in the above kinematic region. It has been claimed that the overall deviation from the SM in B-physics is more than $4\sigma$ when global analyses are performed~\cite{Capdevila:2017bsm, DAmico:2017mtc, Altmannshofer:2017yso} by including other anomalies (the branching ratios of $B\rightarrow K^{(*)}\mu^+\mu^-$~\cite{Aaij:2014pli} and $B_s\rightarrow \phi\mu^+\mu^-$~\cite{Aaij:2015esa}, the angular distribution of decay rate of $B\rightarrow K^{(*)}\mu^+\mu^-$ and $P_5'$ observables~\cite{Aaij:2015oid, Aaij:2014pli, Aaij:2013qta}). If these anomalies are truly due to some physical effects, then lepton flavour universality (LFU) is violated  and new physics beyond the SM is warranted. 

We shall  focus exclusively on anomalies in $\mathcal{R}_{K}$ and $\mathcal{R}_{K^*}$, since the theoretical uncertainty is expected to be small. Economical explanations are involved with four-fermion effective operators, such as $(\bar s \gamma_\mu P_L b)(\bar \ell \gamma^\mu \ell)$, see Ref.~\cite{Alonso:2014csa} for more systematic discussion. More concrete models have also been constructed to generate such an effective operator~\cite{Alonso:2017uky, Sala:2017ihs, Bishara:2017pje, Ellis:2017nrp, Bonilla:2017lsq, Feruglio:2017rjo, Greljo:2017vvb, Alonso:2017bff, Wang:2017mrd, Alok:2017sui, Alok:2017jaf, DiChiara:2017cjq, Kamenik:2017tnu, Cai:2017wry, Ghosh:2017ber, Becirevic:2017jtw, Celis:2017doq}. Various models~\cite{Gauld:2013qja, Buras:2013dea, Altmannshofer:2014cfa, Crivellin:2015mga, Crivellin:2015lwa, Celis:2015ara, Greljo:2015mma, Altmannshofer:2015mqa, Niehoff:2015bfa, Belanger:2015nma, Falkowski:2015zwa, Carmona:2015ena, Chiang:2016qov, Becirevic:2016zri, Boucenna:2016wpr, Megias:2016bde, GarciaGarcia:2016nvr, Ko:2017lzd, Megias:2017ove, Hiller:2014yaa, Gripaios:2014tna, Crivellin:2017zlb, Alonso:2015sja, Sahoo:2016pet, Hu:2016gpe, Chen:2017hir, Ko:2017yrd, Geng:2017svp, Altmannshofer:2017poe} in the literature, including extra $Z'$, lepto-quark and loop-induced mechanisms, were proposed to address similar issues in the past. 

In this paper we focus on $Z'$ models with an extra $U(1)$ gauge symmetry. In the literature, usually just a specific charge assignment is chosen, without noting that many other options could be equally possible. Here, we provide a systematic investigation of general family $U(1)$ gauge symmetry and illustrate how to choose charges consistently to get anomaly-free models without introducing new fermions, except for three right-handed neutrinos. For family universal models, there is only one non-trivial charge assignment, the well-known $B-L$ symmetry. For family non-universal models, however, infinitely many solutions exist as linear combinations of five independent anomaly-free bases. We also show how some models can provide explanations for the anomalies in B-meson decays.

This paper is organized as follows. In Section~\ref{sec:model}, we discuss the consistent conditions for $U(1)'$ charge assignment first, then give an example to show how a realistic model can be constructed to match the observed fermion masses and mixings. In Section~\ref{sec:pheno}, we exemplify one charge assignment in the context of anomalies in B-meson decays. Finally, we give our conclusion.

\begin{table}[t]
 \begin{tabular}{|c|c|c|c|c|}
\hline
 & $SU(3)_{c}$  & $SU(2)_{L}$  & $U(1)_{Y}$  & $U(1)^{\prime}$ \tabularnewline
\hline $Q_{L}^{i}$  & $3$  & $2$  & $+1/3$  & $z_{Q^{i}}=\left( a, b, c \right)$\tabularnewline \hline 
$u_{R}^{i*}$  & $\bar{3}$  & $1$  & $-4/3$  & $z_{u_{i}^{*}}=\left(-a, -b, -c\right)$\tabularnewline \hline 
$d_{R}^{i*}$  & $\bar{3}$  & $1$  & $+2/3$  & $z_{d_{i}^{*}}=\left(-a,-b, -c\right)$\tabularnewline \hline\hline 
$L^{i}$  & $1$  & $2$  & $-1$  & $ z_{L^{i}}= \left(d, e, f\right)$\tabularnewline \hline 
$e_{R}^{i*}$  & $1$ & $1$  & $+2$  &    $ z_{e_{i}^{*}}=\left(-d,-e,-f\right)$\tabularnewline \hline
$\nu_{R}^{i*}$  & $1$  & $1$  & $0$  &  $ z_{\nu_{i}^{*}}=\left(-d,-e,-f\right)$\tabularnewline \hline
\end{tabular}
\caption{An example with anomaly-free realization of $U(1)'$ charges for SM chiral fermions. The last column shows the $U(1)'$ charges for three generations, where $a,b,c,d,e$ and $f$ are arbitrary real numbers. The $U(1)'$ gauge anomaly is canceled in the quark and lepton sectors separately if $c=-(a+b)$ and $f=-(d+e)$, otherwise Eq.~(\ref{eq:abc}) has to  hold. \label{tab:charges}}
\end{table}

\section{Anomaly free Family-Nonuniversal $U(1)'$ Models}\label{sec:model}
In this section, we give some general discussion about anomaly-free conditions for $U(1)'$ models, without introducing extra chiral fermions other than three right-handed neutrinos. We denote the weak doublets and singlets as follows,
$\psi=u,d,e,\nu$,
\[
Q_{L}^{i}=\left(\begin{array}{c}
u_{L}^{i}\\
d_{L}^{i}\end{array}\right),\; L^{i}=\left(\begin{array}{c}
\nu_{L}^{i}\\
e_{L}^{i}\end{array}\right),\;\psi_{L,R}^{i}=P_{L,R}\psi^{i},
\]
with $ P_{L}=(1-\gamma_{5})/2,P_{R}=(1+\gamma_{5})/2$ and $i=1,2,3$ as the family/generation index. The anomaly is proportional to the completely symmetric constant
factor,
\[
D_{\alpha\beta\gamma}\equiv\textrm{tr}\left[\left\{
T_{\alpha},T_{\beta}\right\} T_{\gamma}\right]
\]
$T_{\alpha}$ is the representation of the gauge algebra on the set
of all left-handed fermion and anti-fermion fields, and $``\textrm{tr}"$ stands
for summing over those fermion and anti-fermion species. Note that the $T$s above may or may not be the same since they depend on the referred gauge groups and also the chiral fermions running in the loop of the triangle anomaly-diagram.

The anomaly free conditions for the theory are given by
\begin{eqnarray}
0 & = &
\sum_{i=1}^{3}(2z_{Q_{i}}+z_{u_{i}^{*}}+z_{d_{i}^{*}}),\quad \left[SU(3)^{2}U(1)^{\prime}\right], 
\nonumber\\
0 & = & \sum_{i=1}^{3}(6z_{Q_{i}}+3z_{u_{i}^{*}}+3z_{d_{i}^{*}}+2z_{L_{i}}+z_{e_{i}^{*}}+z_{\nu_{i}^{*}}),\quad \left[\textrm{global } U(1)^{\prime}\right] \nonumber\\
0 & = & \sum_{i=1}^{3}(z_{Q_{i}}^{2}-2z_{u_{i}^{*}}^{2}+z_{d_{i}^{*}}^{2}-z_{L_{i}}^{2}+z_{e_{i}^{*}}^{2}),\quad \left[U(1)^{\prime}{}^{2}U(1)_{Y}\right], \nonumber\\
0 & = &
\sum_{i=1}^{3}(6z_{Q_{i}}^{3}+3z_{u_{i}^{*}}^{3}+3z_{d_{i}^{*}}^{3}+2z_{L_{i}}^{3}+z_{e_{i}^{*}}^{3}+z_{\nu_{i}^{*}}^{3}),\quad \left[U(1)^{\prime}{}^{3}\right].\nonumber\\
0 & = & \sum_{i=1}^{3}(3z_{Q_{i}}+z_{L_{i}}),\quad \left[SU(2)^{2}U(1)^{\prime}\right], \nonumber\\
0 & = & \sum_{i=1}^{3}(\frac{1}{6}z_{Q_{i}}+\frac{4}{3}z_{u_{i}^{*}}+\frac{1}{3}z_{d_{i}^{*}}+\frac{1}{2}z_{L_{i}}+z_{e_{i}^{*}}),\quad \left[U(1)_{Y}^{2}U(1)^{\prime}\right].
\label{eq:anomaly}
\end{eqnarray}
So far, the discussion has been standard and the solution space of the above equations is expected to be large since we have more variables than equations. Interestingly, one can easily check that the first four equations are satisfied automatically if fermions are vector-like under the new $U(1)'$ gauge symmetry, namely 
\begin{equation}
z_{Q_{i}}=-z_{u_{i}^{*}}=-z_{d_{i}^{*}},\; z_{L_{i}}=-z_{e_{i}^{*}}=-z_{\nu_{i}^{*}}.
\end{equation}
With vector-like charge assignment, we only need take care of the last two linear equations, which are actually reduced to just one,
\begin{equation}\label{eq:linear}
3\sum_{i=1}^{3}z_{Q_{i}}=-\sum_{i=1}^{3}z_{L_{i}}.
\end{equation}
This equation is much easier to solve, but could have multiple solutions. For example,
\begin{enumerate}
 \item Family universal model:
 \begin{equation}
	z_{Q}=-z_{L}/3,
 \end{equation}
 which is the unique non-trivial solution, the well-known $B-L$ gauge symmetry. 
 
 \item Family non-universal models:
 \begin{equation}
3\sum_{i=1}^{3}z_{Q_{i}}=-\sum_{i=1}^{3}z_{L_{i}},
 \end{equation}
 where $z_{i}$ are not identical. Since we have six variables but just one constraint, infinitely many solutions exist. For example, we are free to choose just one generation to be charged, the other two as singlets, or any assignments for quark sector with a proper choice of charges for leptons. Some models have been discussed in Refs.~\cite{Liu:2011dh, Xing:2015fdg, Kownacki:2016pmx, Asai:2017ryy}. In general, we can have the charge assignment as in Table~\ref{tab:charges}, where $a,b,c,d,e$ and $f$ are arbitrary real numbers but satisfy
 \begin{equation}\label{eq:abc}
 3(a+b+c)=-(d+e+f).
 \end{equation}
As a special case, we could also imagine that anomalies are canceled separately in the quark and leptons sectors, namely $\sum z_{Q_{i}}=0=\sum z_{L_{i}}$ if $c=-(a+b)$ and $f=-(d+e)$. Such a parametrization includes some well-studied models, such as $a=b=c=0$ and $d=0, e=-f\neq 0$ corresponds to $L_{\mu}-L_{\tau}$, $d=-e\neq 0$ and $f=0$ for $L_{e}-L_{\mu}$, and so on. Note that Eq.~\ref{eq:linear} is linear, so any linear combinations of anomaly-free realizations would also satisfy this equation, like $x(B-L)+y(L_{\mu}-L_{\tau})+z(L_{e}-L_{\mu})+...$ . The solution space for Eq.~(\ref{eq:abc}) is five-dimensional, so we can choose the following five independent solutions as the bases,
\begin{equation}\label{eq:bases}
 L_{e}-L_{\mu},L_{\mu}-L_{\tau}, B_{u}-B_{c}, B_{c}-B_{t}, B-L.  
\end{equation} 

\end{enumerate}

As emphasized above, we are restricting ourselves to extended models with only three additional right-handed neutrinos. If more particles are to be introduced, requirements on the charge assignment should change correspondingly. For example, one could also introduce more SM-singlet Weyl fermions $\chi_j$ with $U(1)'$ charge $X_j$, in cases where SM fermions are vector-like in $U(1)'$, giving
\begin{equation}\label{eq:abcx}
3(a+b+c)+(d+e+f)=0,\; \sum_j X_j=0,\; \sum_j X^3_j = 0.
\end{equation}
Some fermion $\chi_k$ actually could be a dark matter (DM) candidate. For instance, a Majorana mass term $\bar{\chi}_k^c \chi_k$ would be induced after $U(1)'$ symmetry breaking by a SM-singlet scalar $S$ with $U(1)'$ charge $2X_k$, since interactions like $\bar{\chi}_k^c \chi_k S^\dagger$ are allowed. Vector-like $\chi_k$ is another popular scenario for DM where the Dirac mass term $\bar{\chi}_k\chi_k$ is allowed. In both cases, $Z_2$ symmetry can protect the stability of DM. 

To build realistic models with correct SM fermion masses and mixings, we need to introduce some scalar fields $H_i$ to spontaneously break gauge symmetries. The scalar contents would be highly dependent on the charge assignments for these chiral fermions. In the most general cases, for the quark sector we can introduce several Higgs doublets with hypercharge $Y=-1$ and $U(1)'$ charges, $a-b, a-c$ and $b-c$, to make renormalizable Yukawa interactions, giving the desired quark masses and CKM mixing matrix. In  the lepton sector, Higgs doublets with $U(1)'$ charges, $d-e, d-f$ and $e-f$, suffice to give lepton masses and neutrino mixing.

Below, we shall give an example with explicit charge assignment to illustrate how consistent models can be constructed~\cite{Liu:2011dh}. Let us focus on the quark sector first. We shall use the following setup:
\begin{equation}
z_{Q_{i}}=(1,1,-2).
\end{equation}
The above symmetry can be regarded as $3(B_{u}-B_{c})+6(B_{c}-B_{t})$, expanded in the five bases of Eq.~\ref{eq:bases}. Some phenomenologies have been studied first in Ref.~\cite{Liu:2011dh}, and later in Ref.~\cite{Crivellin:2015lwa} along with $L_{\mu}-L_{\tau}$ symmetry in the lepton sector. Here, this model is introduced just for illustration and will be referred to in comparison with the model for the B-anomaly in Section~\ref{sec:pheno}. 

With the above $U(1)'$ charges, a SM Higgs doublet $H_{1}$ with zero $U(1)'$ charge can cause spontaneous electroweak symmetry breaking to generate the masses of all the SM particles, but not correct flavor mixing. To see what happens in the quark sector, we can write the gauge-invariant Yukawa terms as
\begin{eqnarray}
\mathcal{L}_{H_{1}}=\sum_{i,j=1}^{2}\left(y_{ij}^{u}\bar{Q}_{L,i}\tilde{H}_{1}u_{R,j}+y_{ij}^{d}\bar{Q}_{L,i}H_{1}d_{R,j}\right)+y_{33}^{u}\bar{Q}_{L,3}\tilde{H}_{1}u_{R,3}+y_{33}^{d}\bar{Q}_{L,3}H_{1}d_{R,3}+h.c,\label{eq:Y_H1}
\end{eqnarray}
where $y_{ij}^{u,d}$ are the Yukawa couplings. After $H_1$ gets a vacuum expectation value (VEV), the resulting mass matrices for $u$ and $d$ have the following form:
\[
\mathcal{M}_{u,d}^{H_{1}}\sim\left(\begin{array}{ccc}
\times & \times & 0\\
\times & \times & 0\\
0 & 0 & \times\end{array}\right).
\]
This kind of mass matrix cannot give the correct CKM matrix, since the third generation will not mix with the other two. Now if we have two more Higgs doublets, $H_{2}$ with $U(1)'$ charge $-3$ and $H_{3}$ with $+3$, the following Yukawa term are allowed:
\begin{eqnarray}
\mathcal{L}_{H_{2/3}}
& = & y_{13}^{u}\bar{Q}_{L,1}\tilde{H}_{2}u_{R,3}+y_{23}^{u}\bar{Q}_{L,2}\tilde{H}_{2}u_{R,3} + y_{31}^{u}\bar{Q}_{L,3}\tilde{H}_{3}u_{R,1}+y_{32}^{u}\bar{Q}_{L,3}\tilde{H}_{3}u_{R,2}\nonumber \\
& + & y_{13}^{d}\bar{Q}_{L,1}H_{3}d_{R,3}+y_{23}^{d}\bar{Q}_{L,2}H_{3}d_{R,3} + 
y_{31}^{d}\bar{Q}_{L,3}H_{2}d_{R,1}+y_{32}^{d}\bar{Q}_{L,3}H_{2}d_{R,2}+h.c.
\label{eq:Y_H23}
\end{eqnarray}
When both $H_{2/3}$ get VEVs, these terms contribute to the mass matrices with
\[
\mathcal{M}_{u,d}^{H_{2/3}}\sim\left(\begin{array}{ccc}
0 & 0 & \times\\
0 & 0 & \times\\
\times & \times & 0\end{array}\right).\]
Now diagonalizing the total mass matrices, $\mathcal{M}_{u,d}^{H_{1}}+\mathcal{M}_{u,d}^{H_{2/3}}$, would result in three-flavor mixing. Note that we cannot replace $\tilde{H}_3 (H_3)$ with $H_2 (\tilde{H}_2)$ in Eq.~(\ref{eq:Y_H23}) because the $U(1)_Y$ symmetry would forbid that, although only one of them is necessary to give three-flavor mixing. In the case of no $H_3$ or $H_3$ not getting a VEV, the mass matrices are:
\[
\mathcal{M}_{u}^{H_{2}}\sim\left(\begin{array}{ccc}
	0 & 0 & \times\\
	0 & 0 & \times\\
	0 & 0 & 0\end{array}\right), 
\mathcal{M}_{d}^{H_{2}}\sim\left(\begin{array}{ccc}
	0 & 0 & 0\\
	0 & 0 & 0\\
	\times & \times & 0\end{array}\right).
\]
Three-flavor mixing can still arise after diagonalization of $\mathcal{M}_{u,d}^{H_{1}}+\mathcal{M}_{u,d}^{H_{2}}$.
One can easily discuss leptons, since similar physics appears. For example if $z_{L_{i}}=(0,1,-1)$, extra Higgs doublets with charges $\pm 1$ and/or $\pm 2$ would be able to achieve the required lepton masses and mixing.

Gauge bosons will get their masses through the Higgs mechanism. When $H_2$ and $H_3$ get VEVs, the $U(1)'$ gauge symmetry is also broken. If the $U(1)'$ gauge coupling is comparable to the electroweak coupling, the $Z'$ boson is expected to have a mass around the electroweak scale, which is highly constrained. To get a heavy $Z'$ boson, an electroweak singlet scalar $S$ with $U(1)'$ charge $z_{s}$ can be introduced. Then the following vacuum configuration would break the gauge symmetries to $U(1)_{em}$,
\begin{equation}
\langle H_{i}\rangle=\left(
0\; v_{i}/\sqrt{2}\right)^T,\; i= 1,2,3; \qquad \langle S\rangle=v_{s}/\sqrt{2}. 
\label{eq:VEV}
\end{equation}
The kinetic terms for scalars are
\[
\mathcal{L}_{H}=\sum_{i=1}^{3}\left(D^{\mu}H_{i}\right)^{\dagger}\left(D_{\mu}H_{i}\right)+\left(D^{\mu}S\right)^{\dagger}\left(D_{\mu}S\right),
\]
where $D_\mu$ is the covariant derivative. From this Lagrangian, the $W^\pm$ mass can be simply read out, $ g_{2}\sqrt{v_{1}^{2}+v_{2}^{2}+v_{3}^{2}}/2$. Neutral gauge bosons, on the other hand,  are generally mixed, but it is possible to make $Z'$ heavy when $v_s \gg v_i$ such that experimental constraints from $Z-Z'$ mixing can be safely evaded, since the mixing is proportional to $v^2_i/v^2_s$; see Ref.~\cite{Langacker:2008yv} for a general review.

The interaction for $\bar{\psi}\psi Z'$ can be obtained from $gZ_{\mu}^{'}J_{Z^{'}}^{\mu}$, where $g$ is the gauge coupling constant of $U(1)'$ and the current $J_{Z^{'}}^{\mu}$ in the gauge eigenstates is given by
\begin{equation}
J_{Z^{'}}^{\mu}=\sum_{\psi}\sum_{i=1}^{3}\bar{\psi}_{i}\gamma^{\mu}\left[\epsilon_{i}^{\psi_{L}}P_{L}+\epsilon_{i}^{\psi_{R}}P_{R}\right]\psi_{i}\;,\;\psi=u, d, e, \nu. 
\end{equation}
The above $\epsilon_{i}^{\psi_{L/R}}$s are the $U(1)'$ charges $z_{\psi_i}$ for fermions $\psi^{L/R}_i$. Rotating the fermion fields with unitary transformations such that their mass matrices are diagonalized, we get
\begin{eqnarray}\label{eq:f_rotation}
\psi_{R}^{i} & = & \left(V_{\Psi_{R}}\right)_{ij}\Psi_{R}^{j},\;
\psi_{L}^{i}=\left(V_{\Psi_{L}}\right)_{ij}\Psi_{L}^{j},
\end{eqnarray}
where $\Psi=U,D, \bm{e}, \bm{\nu}$ are the mass eigenstates. 
The CKM matrix is given by $ V_{\text{CKM}}=V_{U_{L}}^{\dagger}V_{D_{L}}$ and the neutrino mixing matrix by $V_{\text{PMNS}}=V_{\bm{e}_{L}}^{\dagger}V_{\bm{\nu}_{L}}$. The rotation of fermion fields in Eq.(\ref{eq:f_rotation}) leads to 
\begin{equation}\label{eq:current}
J_{Z^{'}}^{\mu}=\sum_{\Psi=(U,D, \bm{e}, \bm{\nu})}\sum_{i,j=1}^{3}\bar{\Psi}_{i}\gamma^{\mu}\left[ \left(V_{\Psi_{L}}^{\dagger}\epsilon^{\psi}V_{\Psi_{L}}\right)_{ij}P_{L}+\left(V_{\Psi_{R}}^{\dagger}\epsilon^{\psi}V_{\Psi_{R}}\right)_{ij}P_{R}\right]\Psi_{j}.
\end{equation}
We have used $ \epsilon^{\psi}\equiv \epsilon^{\psi_{L}} = \epsilon^{\psi_{R}}$, since we are considering the vector-like charge assignment. One can immediately notice that generally $V^{\dagger}\epsilon V\not\propto I$ if $\epsilon\not\propto I$, namely family non-universal gauge interactions. In our previous examples, we have $ \epsilon^{\psi}\propto\mathrm{diag}\left(1,1,-2\right)$ or $\mathrm{diag}\left(0,1,-1\right)$, and we expect flavor-changing effects to arise. Since only $V_{\text{CKM}}$ or $V_{\text{PMNS}}$ is experimentally measured, the individual matrix $V_{\psi_{L,R}}$ is unknown. Thus the resulting products $V_{\Psi_{L,R}}^{\dagger}\epsilon^{\psi}V_{\Psi_{L,R}}$ are also unknown.

\section{Phenomenologies and anomalies in B-meson decays}\label{sec:pheno}
In this section, we discuss how the above framework can address recent anomalies in B physics. Since left-handed fermions have the same charges as the right-handed ones, we can reparametrize 
\begin{equation}
\epsilon^{\psi}=  z_{\psi_1}I + \mathrm{diag}\left(0,z_{\psi_2}-z_{\psi_1},z_{\psi_3}-z_{\psi_1}\right)\equiv z_{\psi_1}I + \delta \epsilon^{\psi},
\end{equation}
where $z_{U}=(a,b,c), z_{L}=(d,e,f)$, and 
\begin{equation}
B^{\psi_{L,R}}_{ij}\equiv \left(V_{\psi_{L,R}}^{\dagger}\epsilon^{\psi}V_{\psi_{L,R}}\right)_{ij}=z_{\psi_1}\delta_{ij}+(V_{\psi_{L,R}}^{\dagger} \delta \epsilon^{\psi}V_{\psi_{L,R}})_{ij}\equiv z_{\psi_1}\delta_{ij} + \delta B^{\psi_{L,R}}_{ij}.
\end{equation}
Flavor changing processes can happen when $\delta \epsilon^{\psi}\neq 0$ or $\delta B^{\psi_{L,R}}_{ij}\neq 0$.
Note that elements in the matrix $\delta B^{\psi_{L,R}}$ are not necessarily smaller than $ z_{\psi_1}$ for a general setup, since $z_{\psi_1}$ can be zero if fermions in the first generation are $U(1)'$ singlets. 

To illustrate how it affects B meson decay, we exemplify the following anomaly-free charge assignment
\begin{equation}\label{eq:example}
z_{U}=(0,0,1), z_{L}=(0, q_\mu, -3-q_\mu).
\end{equation}
This assignment can be expanded by the bases in Eq.~(\ref{eq:bases}),
\begin{equation} 
(B-L)-(B_{u}-B_{c})-2(B_{c}-B_{t})+(L_{e}-L_{\mu})+(q_\mu+2)(L_{\mu}-L_{\tau}),
\end{equation}
which is a nice example in the sense that it involves all five anomaly-free bases. 
If $q_\mu=-3/2$, the lepton sector has some kind of $L_{\mu}+L_{\tau}$ symmetry. If $|q_\mu|\ll 1$, only the third generation is effectively $U(1)'$-charged. We should emphasize again that it is free to change the above assignment by adding any linear combinations of other anomaly-free solutions. For example, we could use $z'_{U}=(1,1,-1)$ which is just the sum of the above charges with $(1,1,-2)$ mentioned earlier. However, these two models give different signal strengths in experiments, such as LHC dijet events, therefore they are subject to different constraints. 

The $b\rightarrow s$ transitions are usually analyzed in terms of the following effective Hamiltonian
\begin{equation}
\mathcal{H}_{\mathrm{eff}}
=  - \frac{4G_F}{\sqrt{2}} V_{tb} V_{ts}^* \frac{\alpha}{4\pi} \sum_{i}\left(C_i O_i + C_i'O_i'\right) + h.c.
\end{equation}
Here $V$ is the CKM matrix and $\alpha=1/137$ is the fine-structure constant. Note that the coefficients $C_i$ and $C_i'$ are scale-dependent, governed by the renormalization group equation. They are first calculated at high scales and then run to a lower scale, which is usually taken as the bottom quark mass $m_b$ for decay processes. We just list some relevant operators for our later discussions:
\begin{align*}
O_9&=(\bar{s}\gamma_\mu P_L b)(\bar{l}\gamma^\mu l), &&O_9'=(\bar{s}\gamma_\mu P_R b)(\bar{l}\gamma^\mu l), \\
O_{10}&=(\bar{s}\gamma_\mu P_L b)(\bar{l}\gamma^\mu \gamma_5 l), &&O_{10}'=(\bar{s}\gamma_\mu P_R b)(\bar{l}\gamma^\mu \gamma_5 l). 
\end{align*}
In general, all the above operators can be generated. Since anomalies are closely related to $O_9$, we calculate the induced coefficient for $O_9$ by $Z'$-mediated new physics
\begin{equation}
C_9^{\textrm{NP}}\simeq \frac{g^2 \delta B^{D_{L}}_{sb} \left(B^{\bm{e}_{L}}_{\mu\mu}+B^{\bm{e}_{R}}_{\mu\mu}\right)}{2M^2_{Z'}}\biggl{/}\left[\frac{G_F}{\sqrt{2}}  \frac{V_{tb} V_{ts}^*\alpha}{\pi}\right].
\end{equation}
To resolve the anomalies, $C_9^{\textrm{NP}}$ should be around $\simeq -1.1$~\cite{Capdevila:2017bsm, DAmico:2017mtc}, which can be translated into
\begin{equation}\label{eq:para}
\frac{M_{Z'}}{g\sqrt{|\delta B^{D_{L}}_{sb} \left(B^{\bm{e}_{L}}_{\mu\mu}+B^{\bm{e}_{R}}_{\mu\mu}\right)|}}
\simeq 24\TeV.
\end{equation}
The above formula is generally applicable to any non-trivial charge assignment. In some cases, we can simplify it further. For instance, since $B^{\bm{e}_{L/R}}_{\mu\mu}$ are elements in the diagonal, we could expect $B^{\bm{e}_{L}}_{\mu\mu}\sim q_\mu$ if $|q_\mu|\gg|3+q_\mu|$, or no rotation in the charged lepton sector ($V_{\textrm{PMNS}}=V_{\bm{\nu}_{L}}$), and Eq.~(\ref{eq:para}) can be approximated as 
\begin{equation}\label{eq:result}
\frac{M_{Z'}}{g\sqrt{| q_\mu \delta B^{D_{L}}_{sb}|}} \simeq 35\TeV.
\end{equation}
Now with the charge assignment as in Eq.~(\ref{eq:example}), we  explicitly have $\delta B^{D_{L}}_{sb}= (V_{D_{L}}^{\dagger})_{23}(V_{D_{L}})_{33}$. If the CKM matrix comes solely from the rotation of down quarks, we would have $\delta B^{D_{L}}_{sb} =V_{tb} V_{ts}^*$ and 
\begin{equation}\label{eq:parameter}
\frac{M_{Z'}}{g\sqrt{| q_\mu |}} \simeq 7\TeV.
\end{equation}
Other coefficients can also be calculated similarly. Also, if $q_\mu \neq -3$, we would expect new physics effects to show up in $B\rightarrow K^{(*)} \tau^+\tau^-$. Since we mainly focus on $O_9$-related anomalies in B-meson decays to muons, we shall neglect other operators as long as the setup does not violate current limits. For example, we can freely choose $B^{\bm{e}_{L}}_{\mu\mu}=B^{\bm{e}_{R}}_{\mu\mu}$, which results in $C_{10}^{\textrm{NP}} = 0 = C_{10}^{'\textrm{NP}}$.  

$Z'$ may also mediate $B_s-\bar{B}_s$ mixing in the above scenario, since the operator $(\bar{s}\gamma_\mu P_L b)^2$ is inevitably induced, which actually gives the most stringent limit at the moment. Current bounds~\cite{Arnan:2016cpy} can be put on the following quantity:
\begin{equation}\label{eq:bs}
\frac{g^2|\delta B^{D_{L}}_{sb}|^2}{M^2_{Z'}}\lesssim \frac{1}{(300\TeV)^2}, \textrm{ or }
\frac{M_{Z'}}{g}> 12 \TeV \textrm{ for }\delta B^{D_{L}}_{sb} =V_{tb} V_{ts}^*\simeq 0.04 . 
\end{equation}
Comparing with Eq.~(\ref{eq:parameter}), we can safely evade this constraint for $| q_\mu |\gtrsim 3$ and resolve B anomalies at the same time. 

\begin{figure}[t]
	\includegraphics[width=0.5\textwidth,height=0.46\textwidth]{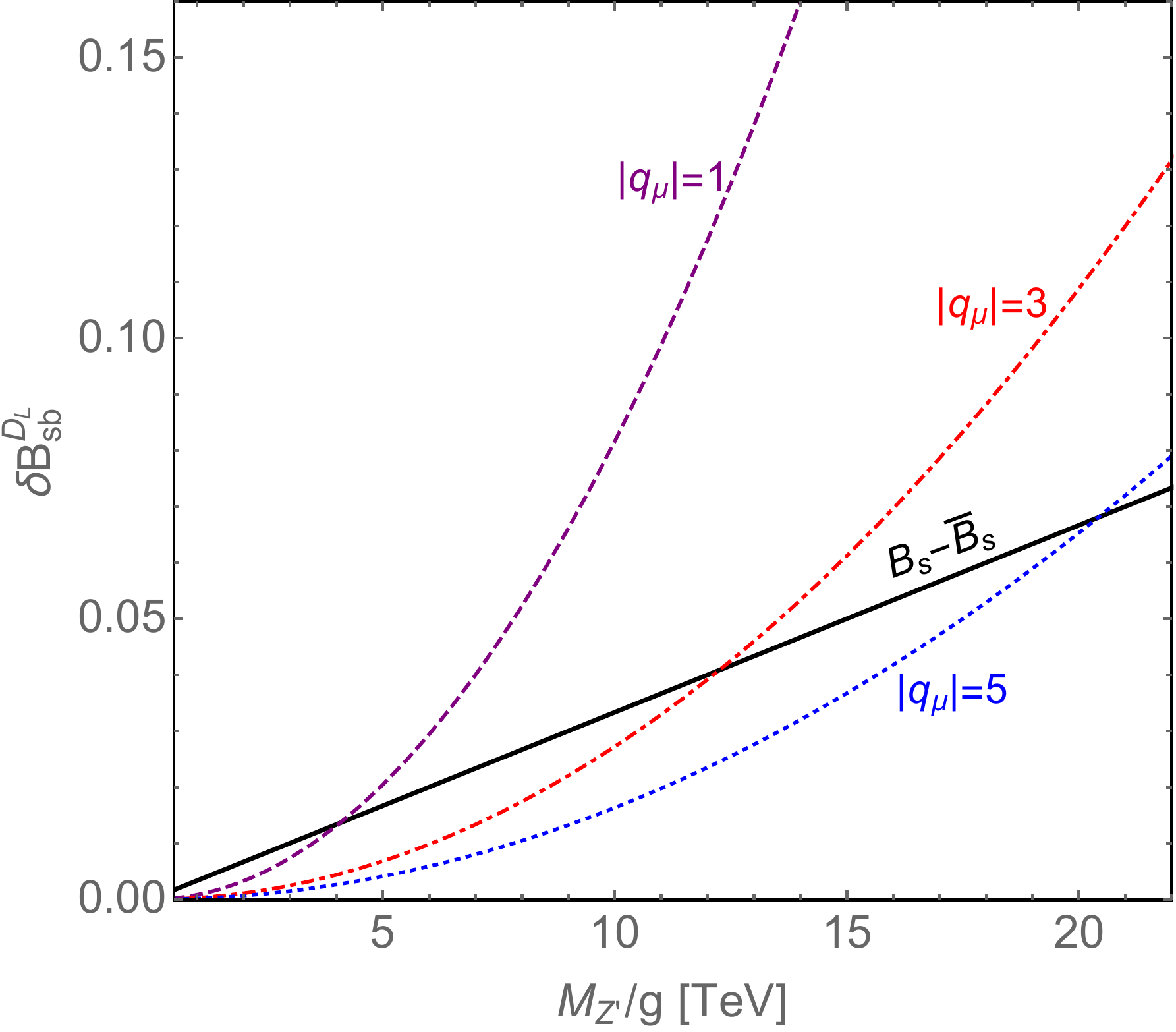}
	\caption{Contours with $C_9^{\textrm{NP}}\simeq -1.1$ in the $M_{Z'}/g$ and $\delta B^{D_{L}}_{sb}$ plane for $|q_\mu|=1,3,5$, shown by dashed purple, dot-dashed red  and dotted blue  lines, respectively. The region above the black line is excluded by $B_s-\bar{B}_s$ mixing.  \label{fig:MzBs}}
\end{figure}

In Fig.~\ref{fig:MzBs}, motivated by B physics anomalies, we plot several contours with $C_9^{\textrm{NP}}\simeq -1.1$ in the $M_{Z'}/g$ and $\delta B^{D_{L}}_{sb}$ plane for $|q_\mu|=1,3$ and $5$. They are shown by the  dashed purple, dot-dashed red  and dotted blue lines, respectively. The region above the black line is excluded by $B_s-\bar{B}_s$ mixing. As expected from Eq.~(\ref{eq:result}), increasing $|q_\mu|$ would allow a larger parameter space. For small $|q_\mu|$, $Z'$ could then be tested by other means. 

\begin{figure}[t]
    \begin{subfigure}{0.48\textwidth}
	\includegraphics[width=\linewidth,height=0.9\linewidth]{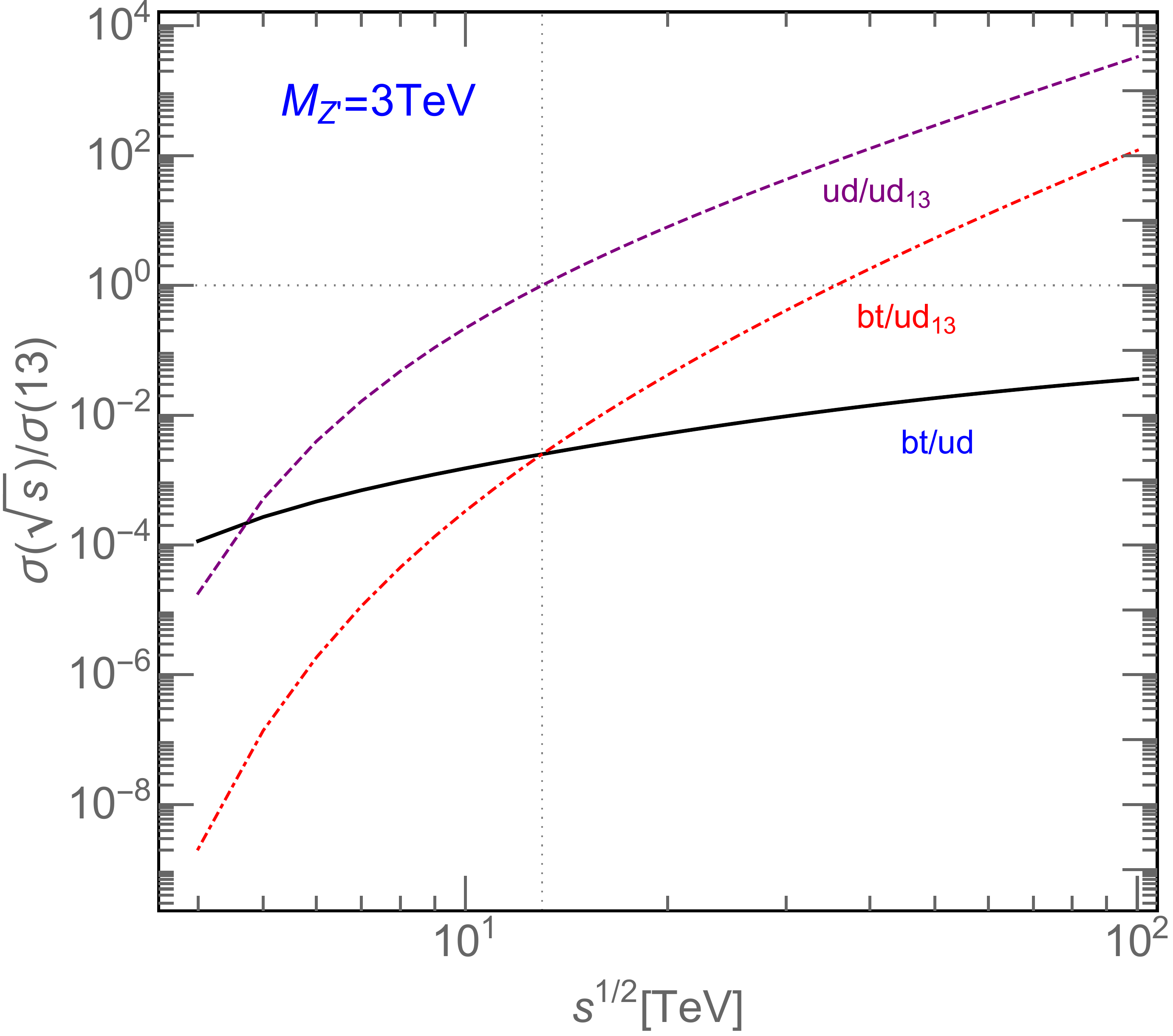}
	\caption{}
    \end{subfigure}	
	\begin{subfigure}{0.48\textwidth}
	\includegraphics[width=\linewidth,height=0.89\linewidth]{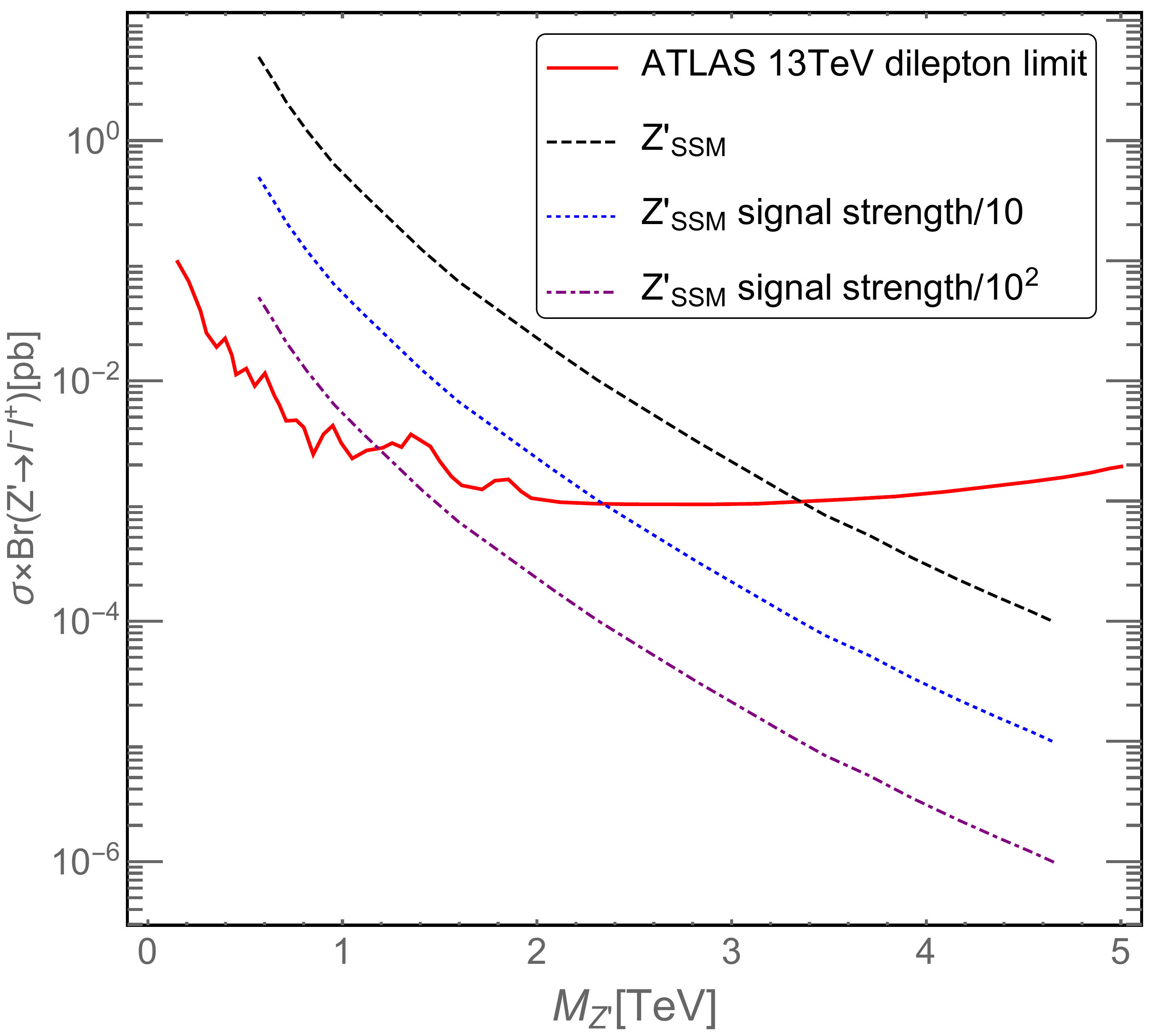}
	\caption{}
	 \end{subfigure}	
	\caption{(a) Various ratios of the production cross section for $Z'$ as functions of energy $\sqrt{s}$, normalized by the cross section when $Z'$ couples to light quarks $u$ and $d$ at LHC with $\sqrt{s}=13\TeV$. The solid black  curve shows the ratios of the contribution from $u$ and $d$ to that from $b$ and $t$. The dotted vertical line indicates $\sqrt{s}=13\TeV$. (b) The mass limit for SSM $Z'$ (dashed black  line) from dilepton searches shifts when the signal strength is reduced by factors of $10$ (dashed blue) and $100$ (dot-dashed purple).  \label{fig:limits}}
\end{figure}

Since $Z'$ couples to both quarks and leptons, dilepton and dijet searches for heavy resonances at colliders can probe $Z'$. The expected signal strength depends on 
\begin{equation}
\sigma \left(f\bar{f}\rightarrow Z'\right)\times Br\left(Z'\rightarrow f'\bar{f}'\right),
\end{equation}
where $\sigma$ is the cross section for $Z'$ production, $f$ and $f'$ are SM fermions, and $Br$ denotes the decay branching ratio. For hadron colliders, we shall integrate the above quantity over the quark parton distribution functions (PDFs) (throughout our calculations, we have used {\tt MMHT2014}~\cite{Harland-Lang:2014zoa} PDFs). In the case of charge assignment for quarks, $(0,0,1)$, hadron colliders such as LHC with energy $\sqrt{s}=13\TeV$ have less discovery potential for $Z'$, since $Z'$ would  couple weakly to the first two generations  through quark mixing only, but strongly to the third generation, which has small PDFs. A future $100\TeV$ hadron collider has a better chance because the production rate is increased thanks to the enhancement of the PDFs of bottom and top quarks. In  Fig.~\ref{fig:limits}(a), we give the ratio of $Z'$ production from bottom and top quarks in our model to that from light quarks if $Z'$ also couples to $u$ and $d$. We have normalized the cross section to a $3\TeV$ $Z'$ at LHC with $\sqrt{s}=13\TeV$. As shown, $Z'$ from the bottom channel is reduced by a factor of $\mathcal{O}(10^3)$ at $\sqrt{s}=13\TeV$ and $\mathcal{O}(10^2)$ at $\sqrt{s}=100\TeV$. Because of that, the limits from hadron collider searches are relaxed dramatically and $M_{Z'}\lesssim 1\TeV$ would still be allowed, which can be inferred from Fig.~\ref{fig:limits}(b), where we show dilepton searches for a Sequential SM (SSM) $Z'$ (SSM $Z'$ is identical to SM $Z$ except for the mass) as the dashed black  line. The region above the solid red  curve is excluded by dilepton searches~\cite{Aaboud:2016cth}. However, if the signal strength is reduced by $10$ or $100$, the exclusion limit would be shifted to $\simeq 2.4 \TeV$ (dashed blue) and $1.2\TeV$ (dot-dashed purple), respectively. Since in the  model discussed, the cross section is lowered by $\mathcal{O}(10^3)$, taking  the branching ratio into account would give $M_{Z'}\gtrsim \mathcal{O}(600\GeV)$, with some dependences on $q_\mu$. Similarly, constraints from dijets are also weakened.

In comparison, the charge assignment $(1,1,-1)$, which is the linear combination of $(1,1,-2)$ in Section~\ref{sec:model} and $(0,0,1)$, will give different results. In such a case, $Z'$ can couple to light quarks and the cross section for production can be sizable. If $g$ is at the same order as the weak coupling, the limit would be similar to the dilepton search at LHC with $\sqrt{s}=13\TeV$ for SSM $Z'$ $M_{Z'}\gtrsim 3.4\TeV$~\cite{Aaboud:2016cth, Khachatryan:2016zqb} and dijet channel $M_{Z'}\gtrsim 3.4\TeV$~\cite{Aaboud:2017yvp} for $g=0.5$. These limits might fluctuate, since the values of the decay branching ratio of $Z'$ would be different from those in $Z'_{SSM}$. In general, direct searches at colliders are complementary to the bound from $B_s-\bar{B}_s$ mixing, Eq.~(\ref{eq:bs}). 

\section{Conclusions}\label{sec:concl}
Motivated by the anomalies in semi-leptonic B-meson decays, we have discussed an explanation in models with general family $U(1)'$ gauge symmetry. We have presented a systematic investigation on how to consistently assign charges to SM chiral fermions with three right-handed neutrinos. If fermions in the standard model are vector-like under this new $U(1)'$ symmetry, their charges in Table~\ref{tab:charges} have to, and only need to, satisfy the condition given in Eq.~(\ref{eq:abc}). Generally, infinitely many anomaly-free family non-universal models exist, as linear combinations of five independent anomaly-free bases. If both bottom quark and muon couple to this new $U(1)$, typically the anomalies in B-meson decays can be explained. 

We have also discussed several other experimental searches for such models, including $Z'$-mediated effects in $B_s-\bar{B}_s$ mixing, dilepton and dijet searches for heavy resonances at colliders. Some viable parameter space has already been probed by these searches. Future searches in colliders and other B-meson decay modes should be able to provide more powerful information on the physical parameters and test different scenarios for $Z'$ charge assignment. 

\vspace*{0.5cm}

\vspace{1 cm}

\centerline{{\bf Acknowledgement}}
YT is grateful to Koichi Hamaguchi, Chengcheng Han and Kazunori Nakayama for enlightening conversations. The work of YT was supported by the Grant-in-Aid for Innovative Areas No.16H06490.

\vspace{20 pt}


\begin{thebibliography}{10}
	
	\bibitem{Aaij:2014ora}
	{\bfseries LHCb} , R.~Aaij {\em et al.}, {\it {Test of lepton universality
			using $B^{+}\rightarrow K^{+}\ell^{+}\ell^{-}$ decays}},
	\href{http://dx.doi.org/10.1103/PhysRevLett.113.151601}{{\em Phys. Rev. Lett.}
		{\bfseries 113} (2014) 151601}
	[\href{http://arxiv.org/abs/1406.6482}{{\ttfamily arXiv:1406.6482}}].
	
	\bibitem{Aaij:2017vbb}
	{\bfseries LHCb} , R.~Aaij {\em et al.},
	{\it {Test of lepton universality with $B^{0} \rightarrow
			K^{*0}\ell^{+}\ell^{-}$ decays}},
	[\href{http://arxiv.org/abs/1705.05802}{{\ttfamily arXiv:1705.05802}}].
	
	\bibitem{RKstar}
	{\bfseries LHCb} , S.~Bifani, {\it Talk by simone bifani for the lhcb
		collaboration, cern, 18/4/2017}, .
	
	\bibitem{Hiller:2003js}
	G.~Hiller and F.~Kruger, {\it {More model-independent analysis of $b \to s$
			processes}},
	\href{http://dx.doi.org/10.1103/PhysRevD.69.074020}{{\em Phys. Rev.} {\bfseries
			D69} (2004) 074020} [\href{http://arxiv.org/abs/hep-ph/0310219}{{\ttfamily
			hep-ph/0310219}}].
	
	\bibitem{Bordone:2016gaq}
	M.~Bordone, G.~Isidori, and A.~Pattori, {\it {On the Standard Model predictions
			for $R_K$ and $R_{K^*}$}},
	\href{http://dx.doi.org/10.1140/epjc/s10052-016-4274-7}{{\em Eur. Phys. J.}
		{\bfseries C76} no.~8, (2016) 440}
	[\href{http://arxiv.org/abs/1605.07633}{{\ttfamily arXiv:1605.07633}}].
	
	\bibitem{Capdevila:2017bsm}
	B.~Capdevila, A.~Crivellin, S.~Descotes-Genon, J.~Matias, and J.~Virto,
	{\it {Patterns of New Physics in $b\to s\ell^+\ell^-$ transitions in the light
			of recent data}},  [\href{http://arxiv.org/abs/1704.05340}{{\ttfamily
			arXiv:1704.05340}}].
	
	\bibitem{DAmico:2017mtc}
	G.~D'Amico, M.~Nardecchia, P.~Panci, F.~Sannino, A.~Strumia, R.~Torre, and
	A.~Urbano, {\it {Flavour anomalies after the $R_{K^*}$ measurement}},
	\href{http://dx.doi.org/10.1007/JHEP09(2017)010}{{\em JHEP} {\bfseries 09}
		(2017) 010} [\href{http://arxiv.org/abs/1704.05438}{{\ttfamily
			arXiv:1704.05438}}].
	
	\bibitem{Altmannshofer:2017yso}
	W.~Altmannshofer, P.~Stangl, and D.~M. Straub, {\it {Interpreting Hints for
			Lepton Flavor Universality Violation}},
	\href{http://dx.doi.org/10.1103/PhysRevD.96.055008}{{\em Phys. Rev.} {\bfseries
			D96} no.~5, (2017) 055008} [\href{http://arxiv.org/abs/1704.05435}{{\ttfamily
			arXiv:1704.05435}}].
	
	\bibitem{Aaij:2014pli}
	{\bfseries LHCb} , R.~Aaij {\em et al.}, {\it {Differential branching fractions
			and isospin asymmetries of $B \to K^{(*)} \mu^+ \mu^-$ decays}},
	\href{http://dx.doi.org/10.1007/JHEP06(2014)133}{{\em JHEP} {\bfseries 06}
		(2014) 133} [\href{http://arxiv.org/abs/1403.8044}{{\ttfamily
			arXiv:1403.8044}}].
	
	\bibitem{Aaij:2015esa}
	{\bfseries LHCb} , R.~Aaij {\em et al.}, {\it {Angular analysis and
			differential branching fraction of the decay $B^0_s\to\phi\mu^+\mu^-$}},
	\href{http://dx.doi.org/10.1007/JHEP09(2015)179}{{\em JHEP} {\bfseries 09}
		(2015) 179} [\href{http://arxiv.org/abs/1506.08777}{{\ttfamily
			arXiv:1506.08777}}].
	
	\bibitem{Aaij:2015oid}
	{\bfseries LHCb} , R.~Aaij {\em et al.}, {\it {Angular analysis of the $B^{0}
			\to K^{*0} \mu^{+} \mu^{-}$ decay using 3 fb$^{-1}$ of integrated
			luminosity}},
	\href{http://dx.doi.org/10.1007/JHEP02(2016)104}{{\em JHEP} {\bfseries 02}
		(2016) 104} [\href{http://arxiv.org/abs/1512.04442}{{\ttfamily
			arXiv:1512.04442}}].
	
	\bibitem{Aaij:2013qta}
	{\bfseries LHCb} , R.~Aaij {\em et al.}, {\it {Measurement of
			Form-Factor-Independent Observables in the Decay $B^{0} \to K^{*0} \mu^+
			\mu^-$}},
	\href{http://dx.doi.org/10.1103/PhysRevLett.111.191801}{{\em Phys. Rev. Lett.}
		{\bfseries 111} (2013) 191801}
	[\href{http://arxiv.org/abs/1308.1707}{{\ttfamily arXiv:1308.1707}}].
	
	\bibitem{Alonso:2014csa}
	R.~Alonso, B.~Grinstein, and J.~Martin~Camalich, {\it {$SU(2)\times U(1)$ gauge
			invariance and the shape of new physics in rare $B$ decays}},
	\href{http://dx.doi.org/10.1103/PhysRevLett.113.241802}{{\em Phys. Rev. Lett.}
		{\bfseries 113} (2014) 241802}
	[\href{http://arxiv.org/abs/1407.7044}{{\ttfamily arXiv:1407.7044}}].
	
	\bibitem{Alonso:2017uky}
	R.~Alonso, P.~Cox, C.~Han, and T.~T. Yanagida, {\it {Flavoured $B-L$ local
			symmetry and anomalous rare $B$ decays}},
	\href{http://dx.doi.org/10.1016/j.physletb.2017.10.027}{{\em Phys. Lett.}
		{\bfseries B774} (2017) 643--648}
	[\href{http://arxiv.org/abs/1705.03858}{{\ttfamily arXiv:1705.03858}}].
	
	\bibitem{Sala:2017ihs}
	F.~Sala and D.~M. Straub,
	{\it {A New Light Particle in B Decays?}},
	[\href{http://arxiv.org/abs/1704.06188}{{\ttfamily arXiv:1704.06188}}].
	
	\bibitem{Bishara:2017pje}
	F.~Bishara, U.~Haisch, and P.~F. Monni, {\it {Regarding light resonance
			interpretations of the B decay anomalies}},
	\href{http://dx.doi.org/10.1103/PhysRevD.96.055002}{{\em Phys. Rev.} {\bfseries
			D96} no.~5, (2017) 055002} [\href{http://arxiv.org/abs/1705.03465}{{\ttfamily
			arXiv:1705.03465}}].
	
	\bibitem{Ellis:2017nrp}
	J.~Ellis, M.~Fairbairn, and P.~Tunney,
	{\it {Anomaly-Free Models for Flavour Anomalies}},
	[\href{http://arxiv.org/abs/1705.03447}{{\ttfamily arXiv:1705.03447}}].
	
	\bibitem{Bonilla:2017lsq}
	C.~Bonilla, T.~Modak, R.~Srivastava, and J.~W.~F. Valle,
	{\it {$U(1)_{B_3-3L_\mu}$ gauge symmetry as the simplest description of $b\to
			s$ anomalies}},  [\href{http://arxiv.org/abs/1705.00915}{{\ttfamily
			arXiv:1705.00915}}].
	
	\bibitem{Feruglio:2017rjo}
	F.~Feruglio, P.~Paradisi, and A.~Pattori, {\it {On the Importance of
			Electroweak Corrections for B Anomalies}},
	\href{http://dx.doi.org/10.1007/JHEP09(2017)061}{{\em JHEP} {\bfseries 09}
		(2017) 061} [\href{http://arxiv.org/abs/1705.00929}{{\ttfamily
			arXiv:1705.00929}}].
	
	\bibitem{Greljo:2017vvb}
	A.~Greljo and D.~Marzocca, {\it {High-$p_T$ dilepton tails and flavor
			physics}},
	\href{http://dx.doi.org/10.1140/epjc/s10052-017-5119-8}{{\em Eur. Phys. J.}
		{\bfseries C77} no.~8, (2017) 548}
	[\href{http://arxiv.org/abs/1704.09015}{{\ttfamily arXiv:1704.09015}}].
	
	\bibitem{Alonso:2017bff}
	R.~Alonso, P.~Cox, C.~Han, and T.~T. Yanagida, {\it {Anomaly-free local
			horizontal symmetry and anomaly-full rare B-decays}},
	\href{http://dx.doi.org/10.1103/PhysRevD.96.071701}{{\em Phys. Rev.} {\bfseries
			D96} no.~7, (2017) 071701} [\href{http://arxiv.org/abs/1704.08158}{{\ttfamily
			arXiv:1704.08158}}].
	
	\bibitem{Wang:2017mrd}
	W.~Wang and S.~Zhao, {\it {Implications of the $R_K$ and $R_{K^*}$ anomalies}},
	\href{http://dx.doi.org/10.1088/1674-1137/42/1/013105}{{\em Chin. Phys.}
		{\bfseries C42} no.~1, (2018) 013105}
	[\href{http://arxiv.org/abs/1704.08168}{{\ttfamily arXiv:1704.08168}}].
	
	\bibitem{Alok:2017sui}
	A.~K. Alok, B.~Bhattacharya, A.~Datta, D.~Kumar, J.~Kumar, and D.~London, {\it
		{New Physics in $b \to s \mu^+ \mu^-$ after the Measurement of $R_{K^*}$}},
	\href{http://dx.doi.org/10.1103/PhysRevD.96.095009}{{\em Phys. Rev.} {\bfseries
			D96} no.~9, (2017) 095009} [\href{http://arxiv.org/abs/1704.07397}{{\ttfamily
			arXiv:1704.07397}}].
	
	\bibitem{Alok:2017jaf}
	A.~K. Alok, D.~Kumar, J.~Kumar, and R.~Sharma,
	{\it {Lepton flavor non-universality in the B-sector: a global analyses of
			various new physics models}},
	[\href{http://arxiv.org/abs/1704.07347}{{\ttfamily arXiv:1704.07347}}].
	
	\bibitem{DiChiara:2017cjq}
	S.~Di~Chiara, A.~Fowlie, S.~Fraser, C.~Marzo, L.~Marzola, M.~Raidal, and
	C.~Spethmann, {\it {Minimal flavor-changing $Z'$ models and muon $g-2$ after
			the $R_{K^*}$ measurement}},
	\href{http://dx.doi.org/10.1016/j.nuclphysb.2017.08.003}{{\em Nucl. Phys.}
		{\bfseries B923} (2017) 245--257}
	[\href{http://arxiv.org/abs/1704.06200}{{\ttfamily arXiv:1704.06200}}].
	
	\bibitem{Kamenik:2017tnu}
	J.~F. Kamenik, Y.~Soreq, and J.~Zupan,
	{\it {Lepton flavor universality violation without new sources of quark flavor
			violation}},  [\href{http://arxiv.org/abs/1704.06005}{{\ttfamily
			arXiv:1704.06005}}].
	
	\bibitem{Cai:2017wry}
	Y.~Cai, J.~Gargalionis, M.~A. Schmidt, and R.~R. Volkas, {\it {Reconsidering
			the One Leptoquark solution: flavor anomalies and neutrino mass}},
	\href{http://dx.doi.org/10.1007/JHEP10(2017)047}{{\em JHEP} {\bfseries 10}
		(2017) 047} [\href{http://arxiv.org/abs/1704.05849}{{\ttfamily
			arXiv:1704.05849}}].
	
	\bibitem{Ghosh:2017ber}
	D.~Ghosh, {\it {Explaining the $R_K$ and $R_{K^*}$ anomalies}},
	\href{http://dx.doi.org/10.1140/epjc/s10052-017-5282-y}{{\em Eur. Phys. J.}
		{\bfseries C77} no.~10, (2017) 694}
	[\href{http://arxiv.org/abs/1704.06240}{{\ttfamily arXiv:1704.06240}}].
	
	\bibitem{Becirevic:2017jtw}
	D.~Bečirević and O.~Sumensari, {\it {A leptoquark model to accommodate
			$R_K^\mathrm{exp} < R_K^\mathrm{SM}$ and $R_{K^\ast}^\mathrm{exp} <
			R_{K^\ast}^\mathrm{SM}$}},
	\href{http://dx.doi.org/10.1007/JHEP08(2017)104}{{\em JHEP} {\bfseries 08}
		(2017) 104} [\href{http://arxiv.org/abs/1704.05835}{{\ttfamily
			arXiv:1704.05835}}].
	
	\bibitem{Celis:2017doq}
	A.~Celis, J.~Fuentes-Martin, A.~Vicente, and J.~Virto, {\it {Gauge-invariant
			implications of the LHCb measurements on lepton-flavor nonuniversality}},
	\href{http://dx.doi.org/10.1103/PhysRevD.96.035026}{{\em Phys. Rev.} {\bfseries
			D96} no.~3, (2017) 035026} [\href{http://arxiv.org/abs/1704.05672}{{\ttfamily
			arXiv:1704.05672}}].
	
	\bibitem{Gauld:2013qja}
	R.~Gauld, F.~Goertz, and U.~Haisch, {\it {An explicit Z'-boson explanation of
			the $B \to K^* \mu^+ \mu^-$ anomaly}},
	\href{http://dx.doi.org/10.1007/JHEP01(2014)069}{{\em JHEP} {\bfseries 01}
		(2014) 069} [\href{http://arxiv.org/abs/1310.1082}{{\ttfamily
			arXiv:1310.1082}}].
	
	\bibitem{Buras:2013dea}
	A.~J. Buras, F.~De~Fazio, and J.~Girrbach, {\it {331 models facing new $b \to
			s\mu^+ \mu^-$ data}},
	\href{http://dx.doi.org/10.1007/JHEP02(2014)112}{{\em JHEP} {\bfseries 02}
		(2014) 112} [\href{http://arxiv.org/abs/1311.6729}{{\ttfamily
			arXiv:1311.6729}}].
	
	\bibitem{Altmannshofer:2014cfa}
	W.~Altmannshofer, S.~Gori, M.~Pospelov, and I.~Yavin, {\it {Quark flavor
			transitions in $L_\mu-L_\tau$ models}},
	\href{http://dx.doi.org/10.1103/PhysRevD.89.095033}{{\em Phys. Rev.} {\bfseries
			D89} (2014) 095033} [\href{http://arxiv.org/abs/1403.1269}{{\ttfamily
			arXiv:1403.1269}}].
	
	\bibitem{Crivellin:2015mga}
	A.~Crivellin, G.~D'Ambrosio, and J.~Heeck, {\it {Explaining
			$h\to\mu^\pm\tau^\mp$, $B\to K^* \mu^+\mu^-$ and $B\to K \mu^+\mu^-/B\to K
			e^+e^-$ in a two-Higgs-doublet model with gauged $L_\mu-L_\tau$}},
	\href{http://dx.doi.org/10.1103/PhysRevLett.114.151801}{{\em Phys. Rev. Lett.}
		{\bfseries 114} (2015) 151801}
	[\href{http://arxiv.org/abs/1501.00993}{{\ttfamily arXiv:1501.00993}}].
	
	\bibitem{Crivellin:2015lwa}
	A.~Crivellin, G.~D'Ambrosio, and J.~Heeck, {\it {Addressing the LHC flavor
			anomalies with horizontal gauge symmetries}},
	\href{http://dx.doi.org/10.1103/PhysRevD.91.075006}{{\em Phys. Rev.} {\bfseries
			D91} no.~7, (2015) 075006} [\href{http://arxiv.org/abs/1503.03477}{{\ttfamily
			arXiv:1503.03477}}].
	
	\bibitem{Celis:2015ara}
	A.~Celis, J.~Fuentes-Martin, M.~Jung, and H.~Serodio, {\it {Family nonuniversal
			Z' models with protected flavor-changing interactions}},
	\href{http://dx.doi.org/10.1103/PhysRevD.92.015007}{{\em Phys. Rev.} {\bfseries
			D92} no.~1, (2015) 015007} [\href{http://arxiv.org/abs/1505.03079}{{\ttfamily
			arXiv:1505.03079}}].
	
	\bibitem{Greljo:2015mma}
	A.~Greljo, G.~Isidori, and D.~Marzocca, {\it {On the breaking of Lepton Flavor
			Universality in B decays}},
	\href{http://dx.doi.org/10.1007/JHEP07(2015)142}{{\em JHEP} {\bfseries 07}
		(2015) 142} [\href{http://arxiv.org/abs/1506.01705}{{\ttfamily
			arXiv:1506.01705}}].
	
	\bibitem{Altmannshofer:2015mqa}
	W.~Altmannshofer and I.~Yavin, {\it {Predictions for lepton flavor universality
			violation in rare B decays in models with gauged $L_\mu - L_\tau$}},
	\href{http://dx.doi.org/10.1103/PhysRevD.92.075022}{{\em Phys. Rev.} {\bfseries
			D92} no.~7, (2015) 075022} [\href{http://arxiv.org/abs/1508.07009}{{\ttfamily
			arXiv:1508.07009}}].
	
	\bibitem{Niehoff:2015bfa}
	C.~Niehoff, P.~Stangl, and D.~M. Straub, {\it {Violation of lepton flavour
			universality in composite Higgs models}},
	\href{http://dx.doi.org/10.1016/j.physletb.2015.05.063}{{\em Phys. Lett.}
		{\bfseries B747} (2015) 182--186}
	[\href{http://arxiv.org/abs/1503.03865}{{\ttfamily arXiv:1503.03865}}].
	
	\bibitem{Belanger:2015nma}
	G.~Bélanger, C.~Delaunay, and S.~Westhoff, {\it {A Dark Matter Relic From Muon
			Anomalies}},
	\href{http://dx.doi.org/10.1103/PhysRevD.92.055021}{{\em Phys. Rev.} {\bfseries
			D92} (2015) 055021} [\href{http://arxiv.org/abs/1507.06660}{{\ttfamily
			arXiv:1507.06660}}].
	
	\bibitem{Falkowski:2015zwa}
	A.~Falkowski, M.~Nardecchia, and R.~Ziegler, {\it {Lepton Flavor
			Non-Universality in B-meson Decays from a U(2) Flavor Model}},
	\href{http://dx.doi.org/10.1007/JHEP11(2015)173}{{\em JHEP} {\bfseries 11}
		(2015) 173} [\href{http://arxiv.org/abs/1509.01249}{{\ttfamily
			arXiv:1509.01249}}].
	
	\bibitem{Carmona:2015ena}
	A.~Carmona and F.~Goertz, {\it {Lepton Flavor and Nonuniversality from Minimal
			Composite Higgs Setups}},
	\href{http://dx.doi.org/10.1103/PhysRevLett.116.251801}{{\em Phys. Rev. Lett.}
		{\bfseries 116} no.~25, (2016) 251801}
	[\href{http://arxiv.org/abs/1510.07658}{{\ttfamily arXiv:1510.07658}}].
	
	\bibitem{Chiang:2016qov}
	C.-W. Chiang, X.-G. He, and G.~Valencia, {\it {Z' model for $b\rightarrow
			sl\bar{l}$ flavor anomalies}},
	\href{http://dx.doi.org/10.1103/PhysRevD.93.074003}{{\em Phys. Rev.} {\bfseries
			D93} no.~7, (2016) 074003} [\href{http://arxiv.org/abs/1601.07328}{{\ttfamily
			arXiv:1601.07328}}].
	
	\bibitem{Becirevic:2016zri}
	D.~Bečirević, O.~Sumensari, and R.~Zukanovich~Funchal, {\it {Lepton flavor
			violation in exclusive $b\rightarrow s$ decays}},
	\href{http://dx.doi.org/10.1140/epjc/s10052-016-3985-0}{{\em Eur. Phys. J.}
		{\bfseries C76} no.~3, (2016) 134}
	[\href{http://arxiv.org/abs/1602.00881}{{\ttfamily arXiv:1602.00881}}].
	
	\bibitem{Boucenna:2016wpr}
	S.~M. Boucenna, A.~Celis, J.~Fuentes-Martin, A.~Vicente, and J.~Virto, {\it
		{Non-abelian gauge extensions for B-decay anomalies}},
	\href{http://dx.doi.org/10.1016/j.physletb.2016.06.067}{{\em Phys. Lett.}
		{\bfseries B760} (2016) 214--219}
	[\href{http://arxiv.org/abs/1604.03088}{{\ttfamily arXiv:1604.03088}}].
	
	\bibitem{Megias:2016bde}
	E.~Megias, G.~Panico, O.~Pujolas, and M.~Quiros, {\it {A Natural origin for the
			LHCb anomalies}},
	\href{http://dx.doi.org/10.1007/JHEP09(2016)118}{{\em JHEP} {\bfseries 09}
		(2016) 118} [\href{http://arxiv.org/abs/1608.02362}{{\ttfamily
			arXiv:1608.02362}}].
	
	\bibitem{GarciaGarcia:2016nvr}
	I.~Garcia~Garcia, {\it {LHCb anomalies from a natural perspective}},
	\href{http://dx.doi.org/10.1007/JHEP03(2017)040}{{\em JHEP} {\bfseries 03}
		(2017) 040} [\href{http://arxiv.org/abs/1611.03507}{{\ttfamily
			arXiv:1611.03507}}].
	
	\bibitem{Ko:2017lzd}
	P.~Ko, Y.~Omura, Y.~Shigekami, and C.~Yu, {\it {LHCb anomaly and B physics in
			flavored $Z'$ models with flavored Higgs doublets}},
	\href{http://dx.doi.org/10.1103/PhysRevD.95.115040}{{\em Phys. Rev.} {\bfseries
			D95} no.~11, (2017) 115040}
	[\href{http://arxiv.org/abs/1702.08666}{{\ttfamily arXiv:1702.08666}}].
	
	\bibitem{Megias:2017ove}
	E.~Megias, M.~Quiros, and L.~Salas, {\it {Lepton-flavor universality violation
			in R$_{K}$ and $ {R}_{D^{{\left(\ast \right)}}} $ from warped space}},
	\href{http://dx.doi.org/10.1007/JHEP07(2017)102}{{\em JHEP} {\bfseries 07}
		(2017) 102} [\href{http://arxiv.org/abs/1703.06019}{{\ttfamily
			arXiv:1703.06019}}].
	
	\bibitem{Hiller:2014yaa}
	G.~Hiller and M.~Schmaltz, {\it {$R_K$ and future $b \to s \ell \ell$ physics
			beyond the standard model opportunities}},
	\href{http://dx.doi.org/10.1103/PhysRevD.90.054014}{{\em Phys. Rev.} {\bfseries
			D90} (2014) 054014} [\href{http://arxiv.org/abs/1408.1627}{{\ttfamily
			arXiv:1408.1627}}].
	
	\bibitem{Gripaios:2014tna}
	B.~Gripaios, M.~Nardecchia, and S.~A. Renner, {\it {Composite leptoquarks and
			anomalies in $B$-meson decays}},
	\href{http://dx.doi.org/10.1007/JHEP05(2015)006}{{\em JHEP} {\bfseries 05}
		(2015) 006} [\href{http://arxiv.org/abs/1412.1791}{{\ttfamily
			arXiv:1412.1791}}].
	
	\bibitem{Crivellin:2017zlb}
	A.~Crivellin, D.~Mueller, and T.~Ota, {\it {Simultaneous explanation of
			R(D$^{(*)}$) and $b\rightarrow s\mu^{+}\mu^{−}$: the last scalar leptoquarks
			standing}},
	\href{http://dx.doi.org/10.1007/JHEP09(2017)040}{{\em JHEP} {\bfseries 09}
		(2017) 040} [\href{http://arxiv.org/abs/1703.09226}{{\ttfamily
			arXiv:1703.09226}}].
	
	\bibitem{Alonso:2015sja}
	R.~Alonso, B.~Grinstein, and J.~Martin~Camalich, {\it {Lepton universality
			violation and lepton flavor conservation in $B$-meson decays}},
	\href{http://dx.doi.org/10.1007/JHEP10(2015)184}{{\em JHEP} {\bfseries 10}
		(2015) 184} [\href{http://arxiv.org/abs/1505.05164}{{\ttfamily
			arXiv:1505.05164}}].
	
	\bibitem{Sahoo:2016pet}
	S.~Sahoo, R.~Mohanta, and A.~K. Giri, {\it {Explaining the $R_{K}$ and
			$R_{D^{(*)}}$ anomalies with vector leptoquarks}},
	\href{http://dx.doi.org/10.1103/PhysRevD.95.035027}{{\em Phys. Rev.} {\bfseries
			D95} no.~3, (2017) 035027} [\href{http://arxiv.org/abs/1609.04367}{{\ttfamily
			arXiv:1609.04367}}].
	
	\bibitem{Hu:2016gpe}
	Q.-Y. Hu, X.-Q. Li, and Y.-D. Yang, {\it {$B^0\to K^{\ast 0}\mu^+\mu^-$ decay
			in the Aligned Two-Higgs-Doublet Model}},
	\href{http://dx.doi.org/10.1140/epjc/s10052-017-4748-2}{{\em Eur. Phys. J.}
		{\bfseries C77} no.~3, (2017) 190}
	[\href{http://arxiv.org/abs/1612.08867}{{\ttfamily arXiv:1612.08867}}].
	
	\bibitem{Chen:2017hir}
	C.-H. Chen, T.~Nomura, and H.~Okada, {\it {Excesses of muon $g-2$,
			$R_{D^{(\ast)}}$, and $R_K$ in a leptoquark model}},
	\href{http://dx.doi.org/10.1016/j.physletb.2017.10.005}{{\em Phys. Lett.}
		{\bfseries B774} (2017) 456--464}
	[\href{http://arxiv.org/abs/1703.03251}{{\ttfamily arXiv:1703.03251}}].
	
	\bibitem{Ko:2017yrd}
	P.~Ko, T.~Nomura, and H.~Okada, {\it {Explaining $B\to K^{(*)}\ell^+ \ell^-$
			anomaly by radiatively induced coupling in $U(1)_{\mu-\tau}$ gauge
			symmetry}},
	\href{http://dx.doi.org/10.1103/PhysRevD.95.111701}{{\em Phys. Rev.} {\bfseries
			D95} no.~11, (2017) 111701}
	[\href{http://arxiv.org/abs/1702.02699}{{\ttfamily arXiv:1702.02699}}].
	
	\bibitem{Geng:2017svp}
	L.-S. Geng, B.~Grinstein, S.~Jäger, J.~Martin~Camalich, X.-L. Ren, and R.-X.
	Shi, {\it {Towards the discovery of new physics with lepton-universality
			ratios of $b\to s\ell\ell$ decays}},
	\href{http://dx.doi.org/10.1103/PhysRevD.96.093006}{{\em Phys. Rev.} {\bfseries
			D96} no.~9, (2017) 093006} [\href{http://arxiv.org/abs/1704.05446}{{\ttfamily
			arXiv:1704.05446}}].
	
	\bibitem{Altmannshofer:2017poe}
	W.~Altmannshofer, P.~Bhupal~Dev, and A.~Soni, {\it {$R_{D^{(*)}}$ anomaly: A
			possible hint for natural supersymmetry with $R$-parity violation}},
	\href{http://dx.doi.org/10.1103/PhysRevD.96.095010}{{\em Phys. Rev.} {\bfseries
			D96} no.~9, (2017) 095010} [\href{http://arxiv.org/abs/1704.06659}{{\ttfamily
			arXiv:1704.06659}}].
	
	\bibitem{Liu:2011dh}
	J.-Y. Liu, Y.~Tang, and Y.-L. Wu, {\it {Searching for $Z^{'}$ Gauge Boson in an
			Anomaly-Free U(1)$'$ Gauge Family Model}},
	\href{http://dx.doi.org/10.1088/0954-3899/39/5/055003}{{\em J. Phys.}
		{\bfseries G39} (2012) 055003}
	[\href{http://arxiv.org/abs/1108.5012}{{\ttfamily arXiv:1108.5012}}].
	
	\bibitem{Xing:2015fdg}
	Z.-z. Xing and Z.-h. Zhao, {\it {A review of $\mu - \tau$ flavor symmetry in
			neutrino physics}},
	\href{http://dx.doi.org/10.1088/0034-4885/79/7/076201}{{\em Rept. Prog. Phys.}
		{\bfseries 79} no.~7, (2016) 076201}
	[\href{http://arxiv.org/abs/1512.04207}{{\ttfamily arXiv:1512.04207}}].
	
	\bibitem{Kownacki:2016pmx}
	C.~Kownacki, E.~Ma, N.~Pollard, and M.~Zakeri, {\it {Generalized Gauge U(1)
			Family Symmetry for Quarks and Leptons}},
	\href{http://dx.doi.org/10.1016/j.physletb.2017.01.013}{{\em Phys. Lett.}
		{\bfseries B766} (2017) 149--152}
	[\href{http://arxiv.org/abs/1611.05017}{{\ttfamily arXiv:1611.05017}}].
	
	\bibitem{Asai:2017ryy}
	K.~Asai, K.~Hamaguchi, and N.~Nagata, {\it {Predictions for the neutrino
			parameters in the minimal gauged U(1)$_{L_\mu-L_\tau}$ model}},
	\href{http://dx.doi.org/10.1140/epjc/s10052-017-5348-x}{{\em Eur. Phys. J.}
		{\bfseries C77} no.~11, (2017) 763}
	[\href{http://arxiv.org/abs/1705.00419}{{\ttfamily arXiv:1705.00419}}].
	
	\bibitem{Langacker:2008yv}
	P.~Langacker, {\it {The Physics of Heavy $Z^\prime$ Gauge Bosons}},
	\href{http://dx.doi.org/10.1103/RevModPhys.81.1199}{{\em Rev. Mod. Phys.}
		{\bfseries 81} (2009) 1199--1228}
	[\href{http://arxiv.org/abs/0801.1345}{{\ttfamily arXiv:0801.1345}}].
	
	\bibitem{Arnan:2016cpy}
	P.~Arnan, L.~Hofer, F.~Mescia, and A.~Crivellin, {\it {Loop effects of heavy
			new scalars and fermions in $b\to s\mu^+\mu^-$}},
	\href{http://dx.doi.org/10.1007/JHEP04(2017)043}{{\em JHEP} {\bfseries 04}
		(2017) 043} [\href{http://arxiv.org/abs/1608.07832}{{\ttfamily
			arXiv:1608.07832}}].
	
	\bibitem{Harland-Lang:2014zoa}
	L.~A. Harland-Lang, A.~D. Martin, P.~Motylinski, and R.~S. Thorne, {\it {Parton
			distributions in the LHC era: MMHT 2014 PDFs}},
	\href{http://dx.doi.org/10.1140/epjc/s10052-015-3397-6}{{\em Eur. Phys. J.}
		{\bfseries C75} no.~5, (2015) 204}
	[\href{http://arxiv.org/abs/1412.3989}{{\ttfamily arXiv:1412.3989}}].
	
	\bibitem{Aaboud:2016cth}
	{\bfseries ATLAS} , M.~Aaboud {\em et al.}, {\it {Search for high-mass new
			phenomena in the dilepton final state using proton-proton collisions at
			$\sqrt{s}=13$ TeV with the ATLAS detector}},
	\href{http://dx.doi.org/10.1016/j.physletb.2016.08.055}{{\em Phys. Lett.}
		{\bfseries B761} (2016) 372--392}
	[\href{http://arxiv.org/abs/1607.03669}{{\ttfamily arXiv:1607.03669}}].
	
	\bibitem{Khachatryan:2016zqb}
	{\bfseries CMS} , V.~Khachatryan {\em et al.}, {\it {Search for narrow
			resonances in dilepton mass spectra in proton-proton collisions at $\sqrt{s}$
			= 13 TeV and combination with 8 TeV data}},
	\href{http://dx.doi.org/10.1016/j.physletb.2017.02.010}{{\em Phys. Lett.}
		{\bfseries B768} (2017) 57--80}
	[\href{http://arxiv.org/abs/1609.05391}{{\ttfamily arXiv:1609.05391}}].
	
	\bibitem{Aaboud:2017yvp}
	{\bfseries ATLAS} , M.~Aaboud {\em et al.}, {\it {Search for new phenomena in
			dijet events using 37 fb$^{-1}$ of $pp$ collision data collected at
			$\sqrt{s}=$13 TeV with the ATLAS detector}},
	\href{http://dx.doi.org/10.1103/PhysRevD.96.052004}{{\em Phys. Rev.} {\bfseries
			D96} no.~5, (2017) 052004} [\href{http://arxiv.org/abs/1703.09127}{{\ttfamily
			arXiv:1703.09127}}].
	
\end{thebibliography}
\providecommand{\href}[2]{#2}\begingroup\raggedright\endgroup

\clearpage
\end{document}